\documentclass[12pt,preprint]{aastex}

\usepackage{epsfig} %,cite}
\def\iso#1#2{\mbox{${}^{#2}{\rm #1}$}}

\def\he#1{\iso{He}{#1}}
\def\li#1{\iso{Li}{#1}}
\def\be#1{\iso{Be}{#1}}
\def\b#1#2{\iso{B}{#1#2}}
\def\bor#1{\iso{B}{#1}}

\def\etal{{\it et al.}~}

\newcommand\beq{\begin{equation}}
\newcommand\eeq{\end{equation}}
\newcommand\beqar{\begin{eqnarray}}
\newcommand\eeqar{\end{eqnarray}}
\def\ga{\mathrel{\mathpalette\fun >}}
\def\fun#1#2{\lower3.6pt\vbox{\baselineskip0pt\lineskip.9pt
  \ialign{$\mathsurround=0pt#1\hfil##\hfil$\crcr#2\crcr\sim\crcr}}}
\begin{document}

%\vspace*{-1.0in}
\slugcomment{
\parbox{1.5in}{
astro-ph/0411728 \\
UMN--TH--2331/04 \\
FTPI--MINN--04/44 \\
November 2004}
} 

\title{Implications of a new temperature scale for halo dwarfs on LiBeB and 
chemical evolution}

\author{Brian D. Fields}
\affil{Center for Theoretical Astrophysics,
Departments of Astronomy and of Physics \\ University of Illinois, Urbana, IL 61801, USA}

\author{Keith A. Olive}
\affil{William I. Fine Theoretical Physics Institute, \\
University of Minnesota, Minneapolis, MN 55455, USA}

\and 

\author{Elisabeth Vangioni-Flam}
\affil{Institut d'Astrophysique, 98 bis Boulevard Arago,
Paris 75014, France}

\begin{abstract}
Big bang nucleosynthesis (BBN) and the cosmic baryon
density from cosmic microwave background
anisotropies together predict a primordial \li7 abundance
a factor of 2--3 higher than that observed in galactic halo dwarf stars.
A recent analysis of \li7 observations in halo stars,
using significantly
higher surface temperature for these stars,
found a higher Li plateau abundance.  
These results go a long way towards resolving the discrepancy
with BBN.
Here, we examine the implications of the higher surface temperatures on the abundances
of Be and B which are thought to have been produced in galactic cosmic-ray nucleosynthesis
by spallation of CNO together with Li (produced in $\alpha + \alpha$ collisions).  While the Be abundance is not overly sensitive to the surface temperature,
the derived B abundances and more importantly the derived 
oxygen abundances are very temperature dependent.
If the new temperature scale is correct,
the implied increased abundances of these elements
poses a serious challenge to models
of galactic cosmic ray nucleosynthesis and galactic chemical evolution.
\end{abstract}


\section{Introduction}

The WMAP-determination of the baryon density 
\citep{wmap,wmap2} allows for very 
accurate predictions of the abundances of the light element isotopes produced in 
big bang nucleosynthesis 
\citep[BBN;][]{cfo3,cva}. While the overall comparison between
these theoretical predictions and the observational determinations of the abundances
of D, \he4, and \li7 are reasonably good,  the theory tends to predict a higher 
\li7 abundance than is observed in the atmospheres of halo dwarf stars.  
This discrepancy is compounded when one takes into account the fact that some
of the observed \li7 has been post-BBN processed in Galactic cosmic-ray nucleosynthesis
as is evidenced by the observations of \li6, Be and B in the same stars.

The WMAP best fit assuming
a varying spectral index is $\Omega_B h^2 = 0.0224 \pm 0.0009$ which is
equivalent to $\eta_{\rm 10,CMB} = 6.14 \pm 0.25$, where $\eta_{10} = 
10^{10} \eta$. This result is
sensitive mostly to WMAP alone but does include CMB
data on smaller angular scales \citep{small,small2}, Lyman $\alpha$ forest data, and
2dF redshift survey data \citep{2df} on large angular scales. 
This result is very similar to the corresponding value obtained
from combining WMAP with SDSS data and other CMB measurements, 
which gives $\Omega_b h^2 = 0.0228^{+0.0010}_{-0.0008}$ \citep{sdss} 
and corresponds to $\eta_{10} = 6.25^{+0.27}_{-0.22}$.  

At the WMAP determined value for $\eta$, D/H = $(2.6 \pm 0.2) \times 10^{-5}$ \citep{cfo3,cyburt, cva}
which can be compared with the observed abundances in quasar absorption systems,
D/H = $(2.8 \pm 0.3) \times 10^{-5}$ \citep{Dref1,Dref2,Dref3,Dref4,Dref5}.
Similarly, good agreement is found when comparing the
predicted \he4 abundance by mass  of $0.2485 \pm 0.0005$ \citep{cfo3,cyburt}
and $0.2479\pm 0.0004$ \citep{cva}
with the observationally determined abundance of $0.2495 \pm 0.0092$ \citep{OSk,OSk2}.
The latter is based on a reanalysis of existing data 
by \citet{iz,iz2}, including effects of 
underlying stellar absorption. Unfortunately, the large observational error precludes
a very discriminatory test using \he4.

In contrast, the BBN predicted value for \li7/H is respectively $(4.3 \pm 0.9) \times 10^{-10}$  \citep{cfo3,cyburt} and $(4.15\pm 0.5) \times 10^{-10}$ \citep{cva}
and is significantly higher than all observational determinations of the \li7 abundance which have ranged in the area of 1-2 $\times 10^{-10}$.  Since the pioneering work of
\citet{spites}, 
who found a value [Li] $\approx 2.1$ corresponding\footnote{[Li] = $\log({\rm Li/H}) + 12$.} to 
 \li7/H $\approx 1.2 \times 10^{-10}$ independent of Fe/H
for [Fe/H] $< -1.3$ and of surface temperature for $T > 5500$ K,  there have been many 
independent observations of Li confirming the existence of a plateau
suggesting a primordial origin for \li7.  The Li abundance extracted from
observations depends heavily on stellar models.  In particular, the final
Li abundance depends sensitively on the assumed surface temperature of the star,
the surface gravity, and the metallicity.  One of the first attempts at
using a systematic set of stellar parameters utilized Balmer lines to obtain surface temperatures
and found a very narrow plateau value of [Li] = $2.224 \pm 0.013$ \citep{mpb}.  The small statistical error
reflects the large number of stars observed and the lack of dispersion in the derived Li abundances.
Shortly thereafter, the infrared flux method (IRFM) was applied to the Li data and new temperatures
were derived yielding a similar value for the Li plateau, [Li] = $2.238 \pm 0.012 \pm 0.05$ \citep{bm},
where the second uncertainty is an estimate of the systematic error. These values correspond to 
\li7/H  = $(1.7 \pm 0.1) \times 10^{-10}$.

The narrowness of the Li plateau was confirmed by \citet{rnb},
who found [Li] = $2.12 \pm 0.02$.  More surprisingly however, their results indicated 
the presence of a slope of Li/H with respect to Fe/H. When numerous corrections for
various systematic effects are accounted for, a Li abundance of $1.2^{+0.7}_{-0.3} \times 10^{-10}$
was derived \citep{rbofn}.

Recent data \citep{bon1} with temperatures
based on H$\alpha$ lines (considered to give systematically high
temperatures) yields \li7/H = $(2.19 \pm 0.28) \times
10^{-10}$. These results are based on a globular cluster sample (NGC 6397).  
This result is consistent with previous Li measurements of the same cluster
which gave  \li7/H = $(1.91 \pm 0.44) \times
10^{-10}$ \citep{pm} and  \li7/H = $(1.69 \pm 0.27) \times
10^{-10}$ \citep{thev}.  A related study (also of globular
cluster stars) gives \li7/H = $(2.29 \pm 0.94) \times10^{-10}$ \citep{bon2}.  
Note that recently, Beryllium has also been observed in this globular cluster, 
the abundance is  $\log({\rm Be/H}) = -12.35 \pm0.2$; 
and is very similar to the abundance found in field halo stars of same metallicity \citep{pas}.
Finally, a set of Li observations based on  6104 \AA\ lines (in contrast to the more common
6707 \AA\ lines) implied a significantly higher \li7 abundance, [Li] $\sim 2.5$ corresponding to
\li7/H $\sim 3 \times 10^{-10}$ \citep{ford}, cautioning however that the weakness of this
line could cause the results to be unreliable.

With the possible exception of the Li abundance determined via the weak 6104 \AA\ lines, 
all observations of Li lead to a serious discrepancy with the BBN predicted value.
This discrepancy assumes that the Li abundance in the stellar
sample reflects the initial abundance at the birth of the star.
However, a possible source of systematic uncertainty comes from the
depletion of Li over the  age of
the Pop II stars. The atmospheric Li abundance will suffer depletion
if the outer layers of the stars have been transported deep enough
into the interior, and/or mixed with material from the hot interior;
this may occur due to convection, rotational mixing, or diffusion.
Standard stellar evolution models
predict Li depletion factors which are very small
($\sigma_{{\rm [Li]}} < $0.05~dex) in very metal-poor turnoff stars
\citep{ddk}. However,  there is no reason to 
believe that such simple models incorporate all effects which lead to
depletion such as rotationally-induced mixing and/or diffusion.
Current estimates for possible depletion factors are in the range
$\sim$~0.2--0.4~dex \citep{dep1,dep2,dep3,dep4}. 
As noted above, the data typically
show negligible intrinsic spread in Li leading to the conclusion
that depletion in these stars is as low as 0.1~dex.

Another potential source for systematic uncertainty
lies in the BBN calculation of the \li7 abundance.  
The prediction for \li7 carries the largest uncertainty
of the four light elements and stems from uncertainties in the nuclear rates.
The effect of changing the yields of certain BBN reactions was
recently considered by \citet{cva}.  For example, 
an increase of either the $\li7(d,n)2\he4$ or
$\be7(d,p)2\he4$ reactions by a factor of 100 would reduce the \li7
abundance by a factor of about 3. 
An experiment has been performed 
at Louvain la Neuve (Belgium) to measure the latter cross section, however
no significant
difference was found compared to previous data 
(C. Angulo, private communication 2004).
The  possibility of systematic errors in the
$\he3(\alpha,\gamma)\be7$ reaction, which is the only important \li7
production channel in BBN, was considered in detail in \citep{cfo4}.  
The absolute value of the cross section
for this key reaction is known relatively poorly both experimentally
and theoretically.  However, the agreement between the standard solar model and
solar neutrino data thus provides additional constraints on variations
in this cross section.  Using the standard solar model of
Bahcall \citep{bah}, and recent solar neutrino data \citep{sno}, one can exclude systematic
variations of the magnitude needed to resolve the BBN \li7
problem at the $\ga 95\%$ CL \citep{cfo4}.  Thus the ``nuclear fix'' to the
\li7 BBN problem is unlikely.

Finally, as noted above, another important source for systematic error
is the assumed  model stellar atmosphere. Its determination
depends on a set of physical parameters and a model-dependent analysis
of a stellar spectrum.  While the dependence on metallicity is typically small,
the surface gravity for hot stars can lead to an underestimate
of up to 0.09 dex if $\log g$ is overestimated by 0.5, though this effect
is negligible in cooler stars.  Typical uncertainties in $\log g$ are
$\pm 0.1 - 0.3$.  The most important source for error is the
surface temperature.  Effective-temperature calibrations for stellar
atmospheres can differ by up to 150--200~K, with higher temperatures
resulting in estimated Li abundances which are higher by $\sim
0.08$~dex per 100~K.  Thus accounting for a difference of 0.5 dex
between BBN and the observations, would require a serious offset of
the stellar parameters.

In this context, the recent work of \citet{mr} is of particular interest. 
While the temperature shift %found in the observations of 
used in the analysis of 
\citet{mr} does not account for the entire BBN discrepancy, 
it does go a long way towards its resolution.  Here, we study the impact of the
new set of temperatures on the abundances of other related elements. 
Indeed, \li7 is also produced by non-thermal processes in the
early Galaxy and afterwards: (i) in Galactic cosmic-ray 
nucleosynthesis (GCRN) and (ii) 
in neutrino spallation taking place during the supernova
explosions.

The evidence for this comes from the related observation of Be and B.
These  elements are observed in the same
halo stars \citep{g2,bk,boes2,mol,pth,prim}, and 
serve as diagnostics of non-primordial lithium production.
Specifically,
the only known source for \be9 and \b10 is GCRN.  While \b11 is also produced 
in GCRN, some \b11 is produced by the $\nu$-process in Type II supernovae \citep{nuproc,opsv,vcfo}.
The theory of GCRN \citep{rfh,mar} is based on either fast protons and $\alpha$s in cosmic rays
spalling C,N, and/or O in the interstellar medium (secondary GCRN) \citep{evf1,fos,fo} or the inverse:
fast CNO breaking up on ISM p's and $\alpha$s 
\citep[primary GCRN;][]{clvf, vroc,hlr,rl,vca}.  
As a consequence, in principle 
the chemical history of Be and B (as well as \li6 which is produced along with \li7 in $\alpha +
\alpha$ collisions) is better traced by the O/H abundance ratio rather than Fe/H. 
Both the abundances of B and O are also highly sensitive to the assumed 
surface temperature in hot halo stars. Thus any significant shift in the temperature,
as reported by Melendez and Ramirez (2004), will have a strong impact on the chemical 
evolution of LiBeB, as well as the stellar and chemical evolution of Oxygen.

In what follows,  we will briefly review the nucleosynthesis of LiBeB in the galaxy
by cosmic rays and neutrinos.  We then apply the new temperature scale
of \citep{mr} to the available BeB and O data.  The consequences of the shifted
abundances are discussed in detail in section 4, where we also discuss
the cosmic-ray nucleosynthesis models needed to describe the data.  
Our conclusions are given in section 5.

\section{LiBeB Nucleosynthesis by Cosmic Rays and Neutrinos}

LiBeB are the ``orphans'' of nucleosynthesis.
Other than \li7, these nuclides are not made in 
observable quantities in the big bang
\citep{tsof}.
Subsequent nucleosynthesis
cannot take place in normal stellar burning
phases, because these fragile nuclei are
readily burned at stellar temperatures.
LiBeB are thus thermodynamically disfavored,
and in fact the bulk of stars are
net sinks of LiBeB.
Thus LiBeB synthesis does not proceed through
thermonuclear processes, as is the case
for essentially all other nuclides.
Instead, LiBeB arise in nonthermal processes,
specifically in two mechanisms: cosmic-ray nucleosynthesis,
and the neutrino process.

\citet{rfh} first recognized that 
cosmic rays are an attractive source for LiBeB production, as
cosmic rays have nonthermal, 
power law spectra, and are typically mildly relativistic.
Cosmic ray propagation thus inevitably leads to
energetic interactions with the interstellar
medium (ISM)\footnote{Note that
structure formation shocks may
generate protogalactic cosmic rays 
as well \citep[e.g.,][]{miniati}
which may also contribute to Li nucleosynthesis
\citep{si,fp}.} 
yielding LiBeB in two ways.
Spallation events 
produce {\em any} LiBeB nucleus
$\ell$ via the fragmentation of heavier
nuclei, 
e.g., $p+{\rm O} \rightarrow \ell + \cdots$.
Note that spallation requires heavy elements (`metals')
already be present in either the cosmic rays or the ISM.
The other class of cosmic-ray--ISM reaction is fusion,
$\alpha+\alpha \rightarrow \li{6,7} + \cdots$
which only produces the lithium isotopes 
\citep{montmerle,sw}.
Note that this does not require any metals,
since helium is always overwhelmingly primordial.

The neutrino process \citep{nuproc,opsv,vcfo}
is another spallation interaction,
one initiated by neutrinos
and taking place in the envelopes of 
core-collapse supernovae.
One of the two important reaction chains
occurs in the carbon shell,
where \bor{11} (and \iso{C}{11})
is produced by the removal of
a nucleon from \iso{C}{12},
e.g., $\nu + \iso{C}{12} \rightarrow \bor{11} + p$;
some of this \bor{11} survives the subsequent passage
of the supernova shock (but
the less-abundant \bor{10} and \be9
do not).
The other key reaction occurs
in the helium shell, where
mass-3 is produced by spallation of \he4,
e.g.,
$\nu + \alpha \rightarrow \iso{H}{3}+p$.
This precedes the shock passage, during which
mass-7 is produced via reactions such as $\iso{H}{3}(\alpha,\gamma)\li7$.

Note that in the neutrino process, the neutrinos
are produced in the collapse and explosion, and the
targets are nuclei made {\em in situ}
during the prior evolutionary phases.
Thus the $\nu$-process does not require
pre-existing metals
in the proto-supernova, and indeed the
\li7 and \bor{11} yields are essentially
independent of the progenitor metallicity
\citep{ww95}.

If cosmic rays and neutrinos are the only
Galactic sources of LiBeB,
then \li6, \be9, \bor{10} are exclusively made by
cosmic rays.
\be9 and \bor{10} require pre-existing metals
in either the cosmic rays or the ISM,
while \li6 does not.
On the other hand, \li7 and \bor{11}
are made both by cosmic rays and the neutrino process.
For \li7, no metals are required for either mechanism,
while for \bor{11} metals are only required for cosmic rays.

It thus follows that
LiBeB history, namely abundance growth with
time and/or metallicity, depends on 
the relative  contributions of cosmic rays  and neutrinos.
This in turn hangs upon the origin of cosmic rays,
details of which are still under debate.
In particular, beryllium (and \bor{10}) nucleosynthesis
by cosmic rays depends on the nature and composition
of the cosmic-ray source material
which is accelerated to high energies.
In the standard scenario,
cosmic rays represent ISM material swept-up and accelerated
by supernova blast waves.
In this picture, cosmic rays have
the same composition as the ISM.
In this case, a metal-dependent species Be or B is
created at a rate
\beq
d{\rm Be}/dt \sim {\rm O} \sigma_{{p{\rm O}}\rightarrow {\rm Be}} \Phi_{p}
\eeq
But the cosmic-ray flux $\Phi_p$ is itself set by
a balance between supernova production and escape,
and thus should
scale with the rate of cosmic ray sources,
i.e., the supernova rate $dN_{\rm SN}/dt$. 
Moreover, since supernovae are the dominant oxygen source,
then $\Phi_p \propto dN_{\rm SN}/dt \propto d{\rm O}/dt$. 
Thus we have
\beq
d{\rm Be}/dt \sim {\rm O} \, d{\rm O}/dt \ \
 \Rightarrow \be9 \propto {\rm O}^2
\eeq
i.e., a quadratic dependence on metallicity,
also known as a ``secondary'' behavior due to the
need for pre-existing metals.

On the other hand, it has been proposed
that the bulk of cosmic rays are not accelerated
in the general ISM, but rather in the rarefied,
metal-rich interiors of superbubbles \citep{clvf,PD}.
These regions are the site of multiple, overlapping supernova remnants,
and thus one expects their hot interiors to 
be dominated by supernova ejecta.
This would mean that the superbubble composition is
enriched in metals, but just as important, the metallicity
would be essentially independent of the global
ISM metallicity.  Thus, any cosmic rays which are
accelerated in superbubble interiors would
have a composition which is both metal-rich and time-independent.
This alters the Be metallicity scaling, since
production is now dominated by the inverse kinematics of
cosmic ray metals on ISM hydrogen, and we have
\beq
d{\rm Be}/dt \sim {\rm O_{CR}} \sigma_{{p{\rm O}}\rightarrow {\rm Be}} \Phi_{p}
  \sim d{\rm O}/dt
\Rightarrow {\be9 \propto O}
\eeq
a linear dependence.
This weaker dependence reflects the metal-independent ``primary'' 
nature of Be production in this case.

Indeed, when Hubble and Keck observations ( and subsequent observations from the 
VLT) of BeB vs Fe
were performed, the necessity of a primary component became apparent. The existence 
of intense fluxes of fast C,O and alpha nuclei in the early Galaxy is now unescapable. 
This low-energy-component (LEC) is physically linked to the superbubble 
scenario (\cite{PD}). In this context, fast nuclei with hard energy spectra at low energies
collide the interstellar medium to produce LiBeB that correspond to primary production.

In general, we expect some combination of 
primary and secondary cosmic-ray nucleosynthesis
to operate (e.g., we expect some supernovae to explode in superbubbles,
and some in the normal ISM).
We thus expect both primary and secondary components of BeB.
It in turn follows that 
in the early Galaxy, 
below some characteristic metallicity scale,
the primary process should dominate, while at later epochs 
the secondary process dominates.
Observationally, we expect a change of slope in 
the Be versus O scaling,
going from linear at low metallicity to
quadratic at high metallicity. 
The change should occur at the metallicity at which each
component contributes equally 
\citep{fovc};
we refer to this as the ``break point''
$Z_{\rm eq}$.
Note that both Li isotopes and \bor{11}
always have primary sources, and thus 
expect these to grow linearly with metallicity
in the early Galaxy regardless of 
the nature of the cosmic ray composition.  Thus, it is only the Be-O trend
which probes the dominant site of cosmic ray origin.

\citet{fovc}
analyzed BeB data extensively to search for these
predicted trends, and to attempt to quantify
the primary to secondary transition.
These authors
found that the
break point metallicity depends on 
the temperature scale used to derive BeB and O abundances,
and the dataset used.
Results spanned a range 
from $[{\rm O/H}]_{\rm eq} = - 1.94$
to
$[{\rm O/H}]_{\rm eq} = - 1.35$. Note that recent measurements \citep{petal,petal2}
of \be{9} at very low metallicity giving a high abundance of this element
should move $Z_{\rm eq}$ towards higher values.
As we will see, the new temperature scale
has an even more drastic effect on our
picture of LiBeB evolution.

\section{A new temperature scale}

The surface temperature is one of the key stellar input parameters needed to 
extract abundances from spectral observations. As abundances can be very sensitive to
the assumed temperature, the lack of a standard to obtain 
temperatures is a large source of systematic uncertainty in 
stellar abundances. As the surface temperature of hot halo stars 
can not be determined directly, indirect methods must be employed. 
These include the use of Balmer lines 
H$\alpha$ through H$\delta$ \citep{axer,afg}, the IRFM \citep{alonso},
and H$\alpha$ lines alone \citep{grat}. 

In order to understand the origin of the LiBeB isotopes, their abundance trends
with metallicity (O/H or Fe/H) must be obtained with some degree of certainty.  Unfortunately,
because of the lack of a standard for stellar input parameters, the literature 
data on the abundance data for these elements must be treated with care.
Previously \citep{fovc}, we recompiled the data for Be and B assuming a common
set of temperatures, surface gravities and metallicities based on the Balmer line method,
the IRFM, and a recalibration of the IRFM \citep{i2}. A similar recalibration was performed for O/H
which is also very sensitive to the stellar input parameters. Among the goals of that work was
to test specifically for the primary vs secondary production of Be and B. A  similar study was undertaken
by \citet{king}.

Recently, using new spectroscopic and photometric surveys,   \citet{mr} have determined an improved IRFM for a large number of stars which are of interest in studying
LiBeB abundances.  The new temperature scale (which we will refer to as MR)
was applied directly to the question
of \li7 abundances and compared with previous results \citep{rnb}.  The new results \citep{mr}
find virtually no scatter in the \li7 abundances supporting the existence of 
a primordial plateau for \li7 with a higher abundance, [Li] $ = 2.37 \pm 0.06$ corresponding to
\li7/H = $(2.34 \pm 0.32) \times 10^{-10}$. While this value eases the discrepancy between
the BBN prediction, it is still about a factor of 2 smaller.
Unlike the data of \citet{rnb}, the recent sample shows no trend 
of \li7 with respect to either temperature or Fe/H. We will discuss the implications of 
this lack of trend in the context of GCRN below.  The overall increase in 
the \li7 abundance was argued to be directly related to the systematically higher 
temperatures assumed.  Indeed, the slope of [Li] vs. [Fe/H] was erased
because the improved IRFM temperatures are systematically higher at low metallicity than
the temperatures used in \citet{rnb}. 

We expect that a new temperature scale will have a significant impact on the
abundances of other elements.  While the Be abundance is not overly sensitive to the
assumed surface temperature,  the abundances of B and O are very sensitive to
the choice of a temperature scale.  For example, a change in temperature of 
100 K leads approximately to a change in the B abundance of 0.12 dex.  Thus
a 200 K increase in the surface temperature would increase a derived boron
abundance by nearly a factor of 2.  Similarly, a 100K change in temperature
leads to roughly a 0.2 dex change in the oxygen abundance. In contrast, 
the beryllium abundance changes by only 0.02 - 0.04 dex for a 100K change in temperature.
This is particularly important since the predicted ratios of elements such as B to Be
can be traced back directly to their cross-sections and thus are largely
model independent.

\begin{figure}[h]
\begin{center}
\epsfig{file=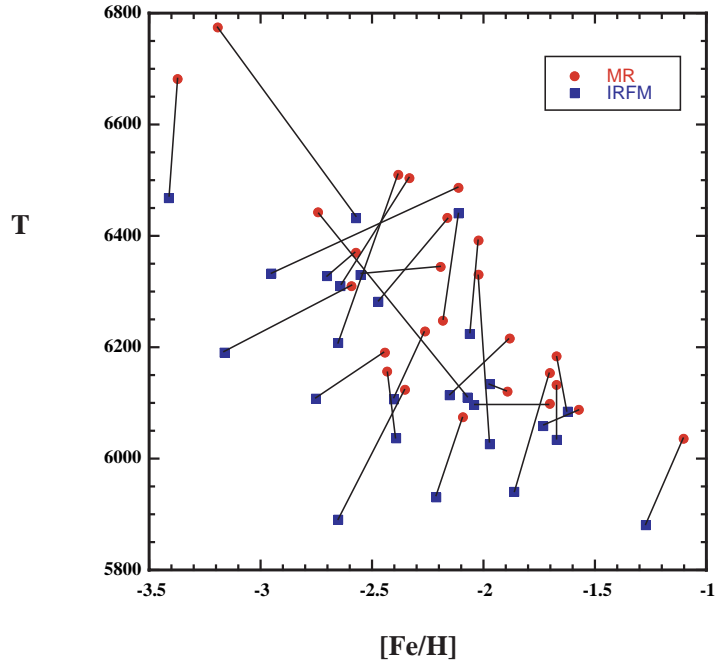, height=3.5in}
\end{center}
\caption{The surface temperature plotted versus [Fe/H] for 25 halo stars.
IRFM parameters from \citet{alonso} are shown as filled squares and parameters from \citet{mr}
are shown as filled circles.  Lines are shown connecting the parameter choices for each star.}
\label{tvsfe}
\end{figure}

However, as noted above, the abundance
of the LiBeB elements is not our only concern.  In order to model the data in the 
context of GCRN, we must have consistent metallicity tracers such as Fe/H and O/H.
Previously \citep{fovc}, we noted a specific troublesome example: that of the star BD $3^\circ$ 740.
{}From the Balmer line method \citep{axer,afg}, the star has derived parameters
($T_{\rm eff}, \log g$, [Fe/H]) = (6264, 3.72, -2.36).
The beryllium and oxygen abundances for this star 
are \citep{bk,boes2} [Be/H] = -13.36, [B/H] = -10.43, and [O/H] = -1.74 when adjusted for
these stellar parameters.
In contrast, the stellar parameters from the IRFM \citep{alonso}
are (6110,3.73,-2.01) with corresponding Be, B and O abundances of
-13.44,  -10.24, and -2.05.  With a recalibrated IRFM, the following parameters are found
\citep{gletal}, (6295,4.00, -3.00).
For these choices, we have [Be/H] = -13.25, [B/H] = -10.98, and [O/H] = -1.90.
Based on the new work \citep{mr}, using (6443, 3.76, -2.68), we
find [Be/H] = -13.25, [B/H] = -10.55, and [O/H] = -1.76.
Notice the extremely large range in assumed metallicities and the
difference in the temperatures.  
These differences are exemplified in Figure \ref{tvsfe}
where we plot the temperatures and metallicities of 25 stars
with parameters taken from the IRFM \citep{alonso} and the work of \citet{mr}.
One can clearly see the systematically higher temperatures
associated with the MR scale.  
Clearly this type of systematic uncertainty makes it very difficult
to draw firm conclusions on the evolution of the LiBeB elements.

In what follows, we will make use of the new temperature scale \citep{mr}
and rederive the abundance of Be, B, and O for those stars appearing in the
compilation of \citet{mr}.  We then compare the new abundances with 
models of GCRN.  

\section{Results}

We present analysis of LiBeB, and O/Fe trends in light
of the new halo star temperature scale.
We first present results of model-independent fitting
procedures, then we compare the data to
the results of  GCRN models.

\subsection{Data Fits}

As discussed in section 2, of key importance towards the understanding of
the origin of the LiBeB elements, is their evolution with respect to metallicity.
In particular, we will be interested in the trends of 
Be/H, B/H, B/Be vs. Fe/H and O/H.% as well as O/Fe vs. Fe/H. 
 Unfortunately, we will be able to use only a subset of the available 
data on Be, B, and O for which we have stellar input parameters derived in \citet{mr}.
This allows us to consider the abundances of Be in 14 stars, B in only 3 stars, and 
O/H in 13 stars.  

\begin{figure}[ht]
\begin{center}
\epsfig{file=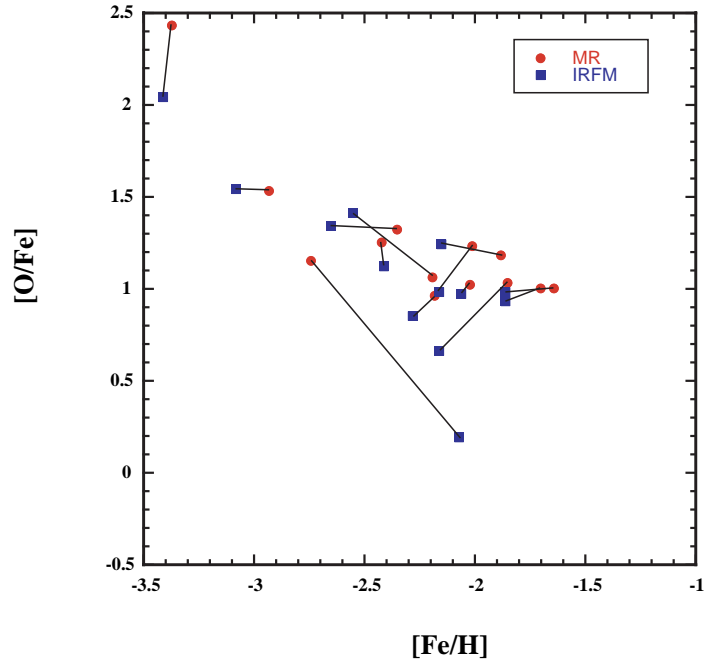, height=3.5in}
\end{center}
\caption{The abundance ratio [O/Fe] versus [Fe/H] for 13 halo stars.
IRFM parameters from \citet{alonso} are shown as filled squares and parameters from \citet{mr}
are shown as filled circles.  Lines are shown connecting the abundances for each star.}
\label{ofe}
\end{figure}

The evolution of the oxygen abundance, particularly O/Fe, has been 
the subject of considerable controversy \citep{mel, is,boes,fk,ietal,ngeg}, 
which we do not intend to partake in here.  
We do note that the oxygen abundances based on the 
MR temperature scale  \citep{mr}
are on the average 0.25 dex higher than the oxygen abundances derived using the IRFM
as tabulated in  \citet{alonso}. 
The affect on the oxygen abundances is shown in Fig. \ref{ofe}.
 While the derived value of [O/H] is not very sensitive to 
[Fe/H], clearly the value of [O/Fe] is.  Thus in Fig. \ref{ofe}, the increase in
O/H is sometimes masked by an increase in [Fe/H].

The  higher
O/H abundances further compounds the problem associated with
high O/Fe ratios at low Fe/H in chemical evolutionary
models with standard yields of type II supernovae at low metallicity. 
In stellar models employing standard supernova yields, one obtains
[O/Fe] abundances which are typically flat at low metallicity
with values of [O/Fe] $\approx$ 0.5 - 0.6; however, the precise value depends
on the initial masses and metallicities of stars \citep{ww95}.  
Oxygen abundances 
based on OH lines have shown rising [O/Fe] with decreasing [Fe/H]
which could possibly be explained by a diminished Fe yield at low metallicity.
The subset of points considered here, has a relatively large (negative) slope
for [O/Fe] vs [Fe/H].  While the full set of IRFM data for 27 stars considered in
\citet{fovc} gave a slope of $-0.44 \pm 0.10$, the subset of 13 stars considered here
(with MR parameters)  yield a much large slope for  [O/Fe] vs [Fe/H] ($-0.89 \pm 0.22$)
using IRFM parameters.  This is largely due to the absence of stars with metallicity
[Fe/H] = -1.5 to -1.0 which generally soften the derived slope.
The analysis of \citet{king} found a lower but non-zero slope
of $-0.25$. In contrast, 
the newly analyzed data considered here shows a lower slope
for [O/Fe] vs [Fe/H] ($-0.66 \pm 0.24$).
The increased oxygen abundances in absolute terms will present a challenge to 
both supernova and chemical evolution
models.

\begin{figure}[ht]
\begin{center}
\epsfig{file=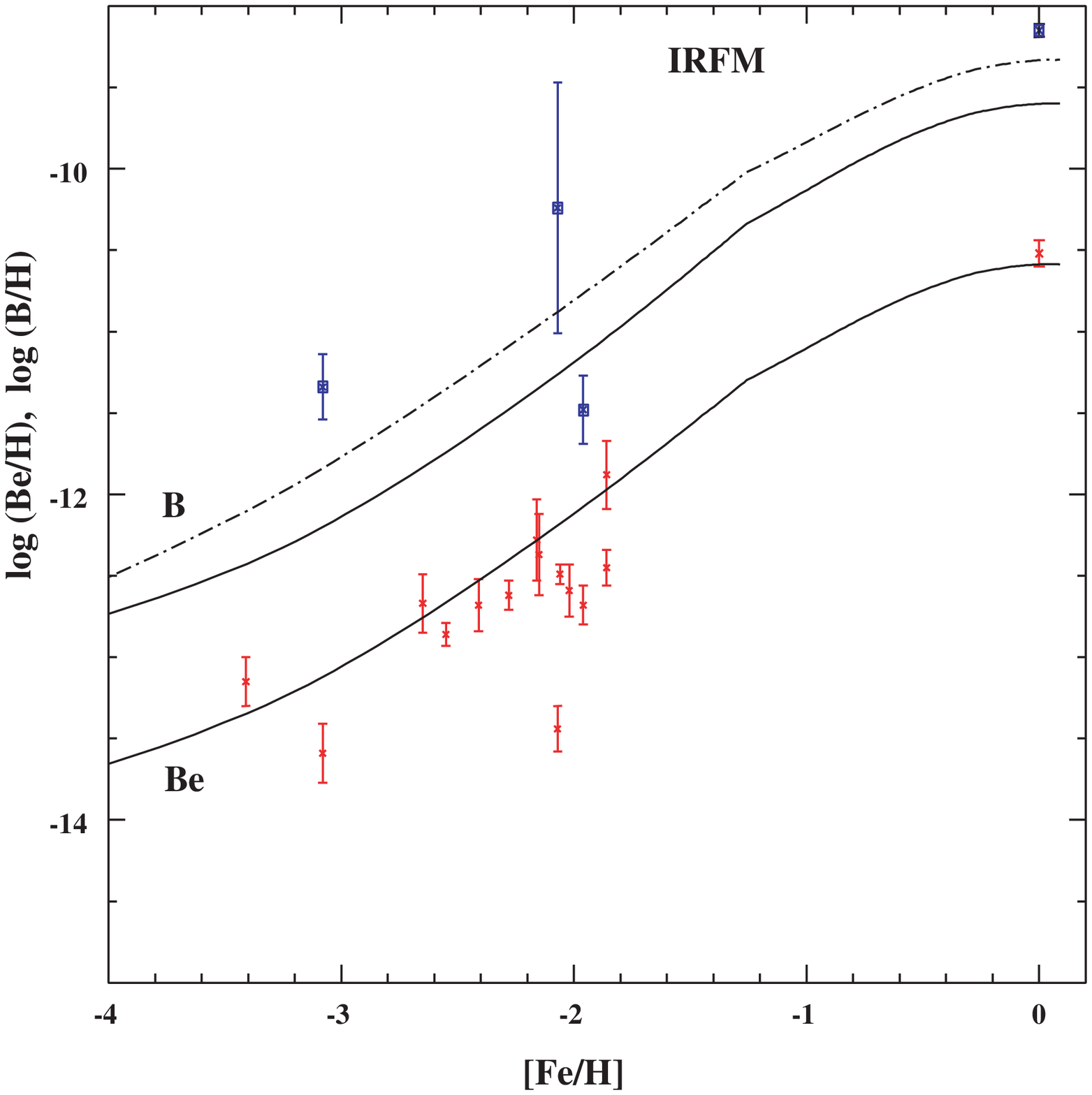, height=3.0in}
\epsfig{file=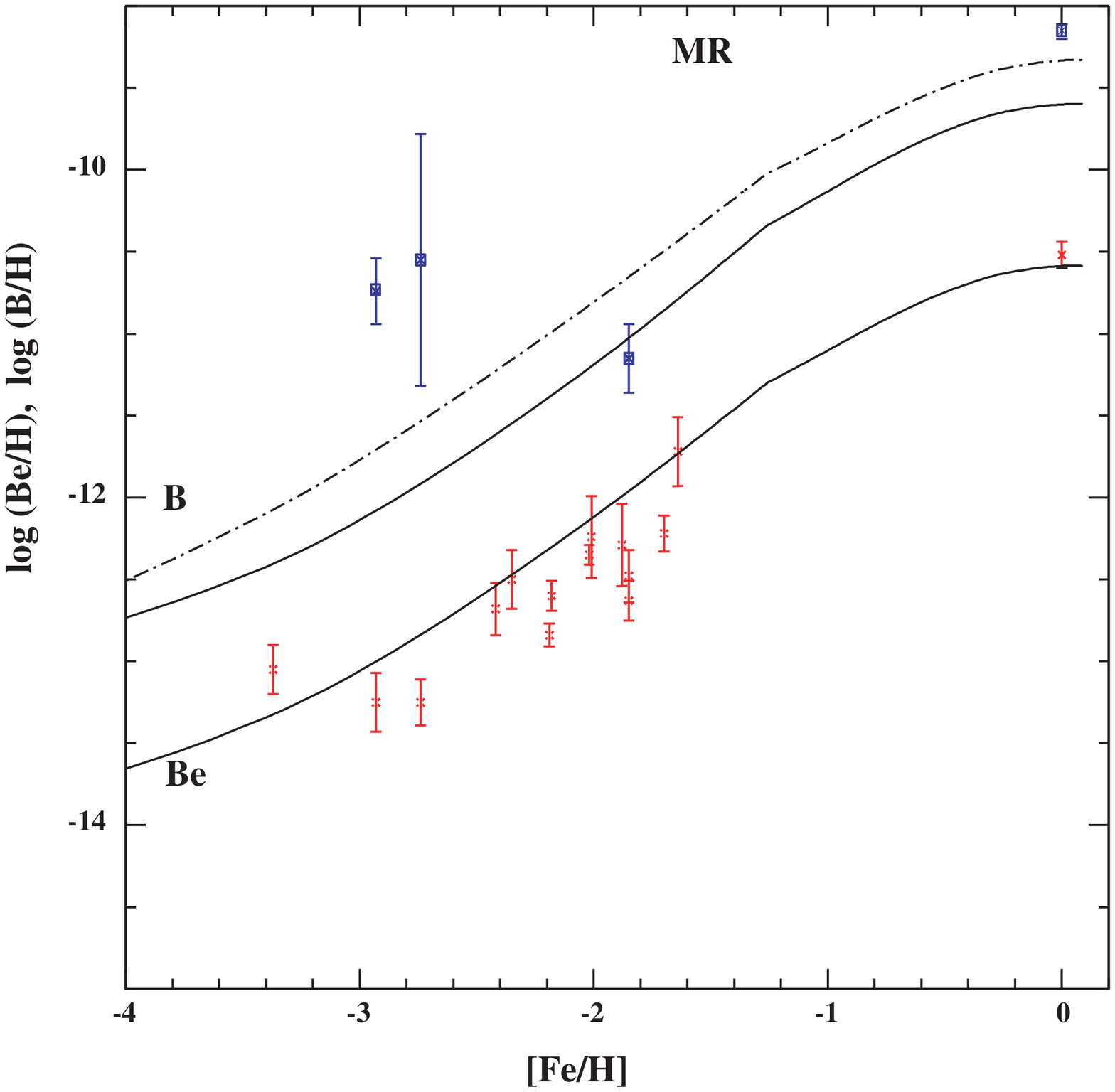, height=3.0in}
\end{center}
\caption{The abundance evolution of log(Be/H) and log(B/H) with respect to  [Fe/H] for 14 halo stars.
In the left panel, the data points are based on the IRFM parameters from \citet{alonso}. 
In the right panel, abundances are based on MR parameters from \citet{mr}.
The square (blue) points correspond to the boron data
and the crosses (red points) to the beryllium data. 
The solid curves  represent the evolution of these two elements
taking into account the secondary and primary RCG processes (see text). The dashed curve shows
the  boron evolution including the neutrino process .}
\label{beba}
\end{figure}

As discussed above, the
Beryllium  abundances are not overly affected by the change in the temperature scale.
In Fig. \ref{beba},   we plot the Be data vs [Fe/H].  In the left panel, we
show the data based on the IRFM, and on the right the data based on MR
for the 14 stars with both MR parameters and measured Be abundances.  
There is a systematic shift in the data towards higher metallicity (which is also seen
in Fig. \ref{tvsfe} by comparing the positively sloped connecting lines to the negatively sloped ones).
We also note that at low metallicity, the MR data seem to show a plateau at very low metallicity 
([Fe/H] $< -2.5$) at [Be/H] = -13.0 -- -13.2, though the poor statistics does not allow
one to draw a definite conclusion.  Furthermore, the plateau is more apparent due to the
shift in one point (corresponding to BD -13 3442) which shows an increase in Be 
of 0.35 dex due in part to the increased temperature (6159 to 6484) and to the increase in 
surface gravity (3.5 to 3.98). 

Using all of the available data based on 25 stars, the IRFM scale implied a slope for Be vs Fe of 1.19 $\pm$ 0.10 \citep{fovc} characteristic of primary GCRN with respect to [Fe/H].  
When we limit the IRFM sample to the 14 stars with MR parameters, the slope is reduced to 0.83 $\pm$ 0.15 and is similar to the slope for the same 14 points using MR parameters (0.92 $\pm$ 0.18).  While the average Be abundance was shifted up by 0.12 dex, the average increase
in Fe/H leaves the slope more or less unchanged.

Unfortunately only 3 of the 15 low metallicity 
stars with observed B are included in the MR sample.  
These are also plotted in both Figs. \ref{beba}.
In addition to being very sensitive to temperature, the derived boron abundance
is also very sensitive to metallicity.  The 3 stars show changes in [B/H] of
0.33, -0.31, and 0.60 with shifts in [Fe/H] of 0.11, -0.67, and 0.15 respectively.
When corrected for the shift in metallicity and surface gravity, the average 
shift in B/H is about 0.3 dex due to the new temperature scale.  
The same 3 points show shift in O/H of 0.68, 0.29, and 0.14.
The relatively small change in the 3rd point  (BD -13 3442 discussed above) is due to
a cancellation in the shift due to temperature and that from the large change in surface gravity.
Because we use B abundances from only three
stars, it is not meaningful at this time to perform fits of B with respect to either Fe/H or O/H.

\subsection{Cosmic-Ray Nucleosynthesis Models}

We now confront GCRN models with the new BeB--OFe trends.
Because the Be data most certainly show a linear trend with
respect to Fe/H at low metallicity we 
illustrate the impact of the new data in two ways.
To begin with we
present a model  \citep{vroc} which includes a primary component
of metal-rich particles with a composition of
$60-100 M_\odot$ stars, presumably found in superbubble interiors,
which dominate LiBeB synthesis in the early Galaxy,
in addition the standard GCR component.
Below we also consider 
a standard GRCN model, with a non-constant  [O/Fe] vs [Fe/H] relation.
In both cases, the GCRN models are accompanied by
the $\nu$-process synthesis of \li7 and \bor{11}.

In Fig. \ref{beba}, we
show a model curve for the GCRN evolution of Be with respect to
[Fe/H] which includes the LEC. 
This model combines secondary and primary processes as explained above, with the same
prescriptions used in \citep{vroc}. The dashed line shows the effect of adding the neutrino-process
contribution to boron. As one can see, the model does a reasonably good job describing the Be
data using either the IRFM or MR parameters, since Be is the least affected by the choice of 
temperature scale.  In contrast, the B data has shifted upwards, but 
it is difficult to draw firm conclusions based on three points.  If the trend
persists with the remaining B data, we would expect large B/Be ratios at low metallicities.
This is a trend not predicted in the CGRN + LEC model.

\begin{figure}[ht]
\begin{center}
\epsfig{file=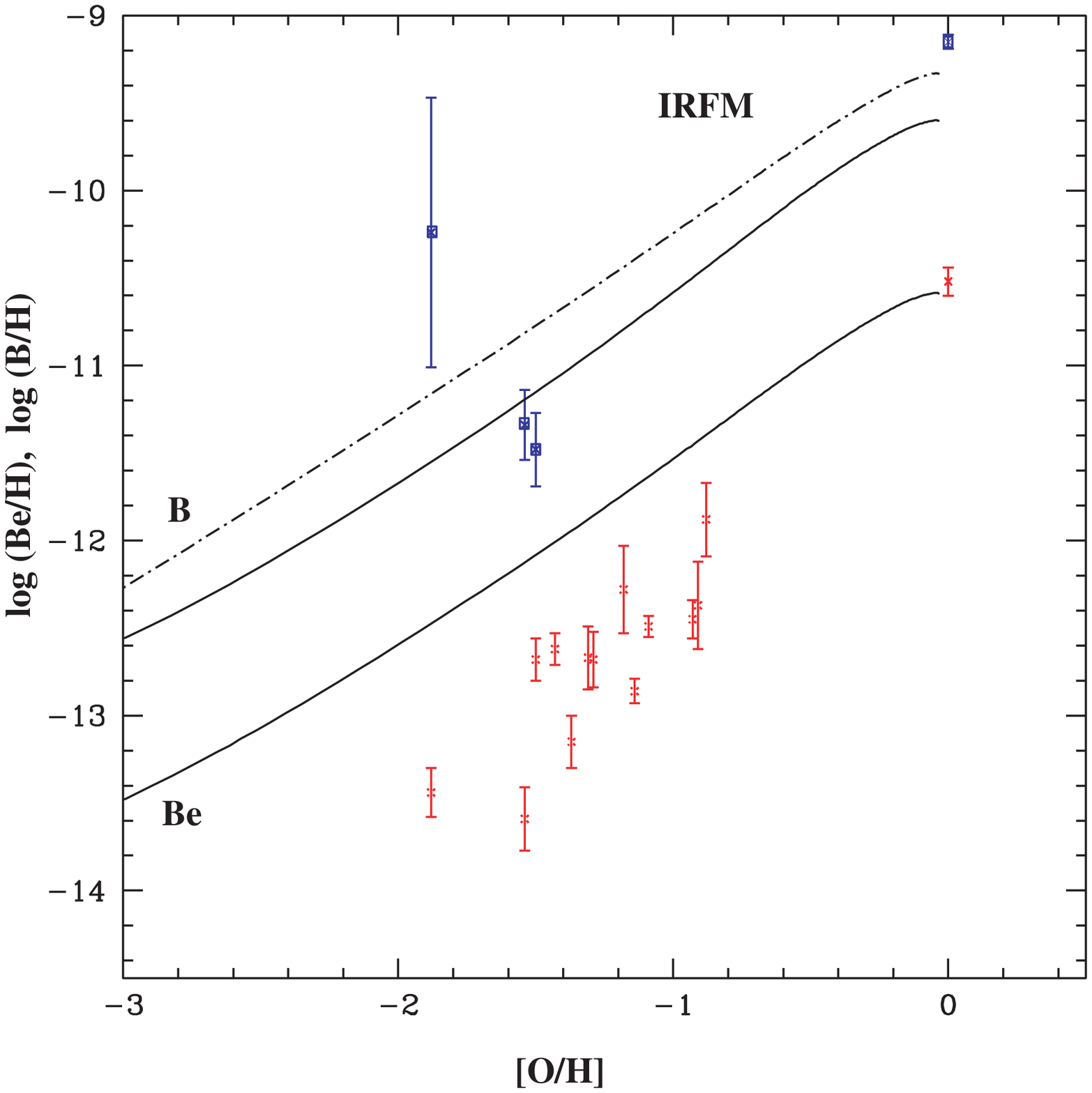, height=3.0in}
\epsfig{file=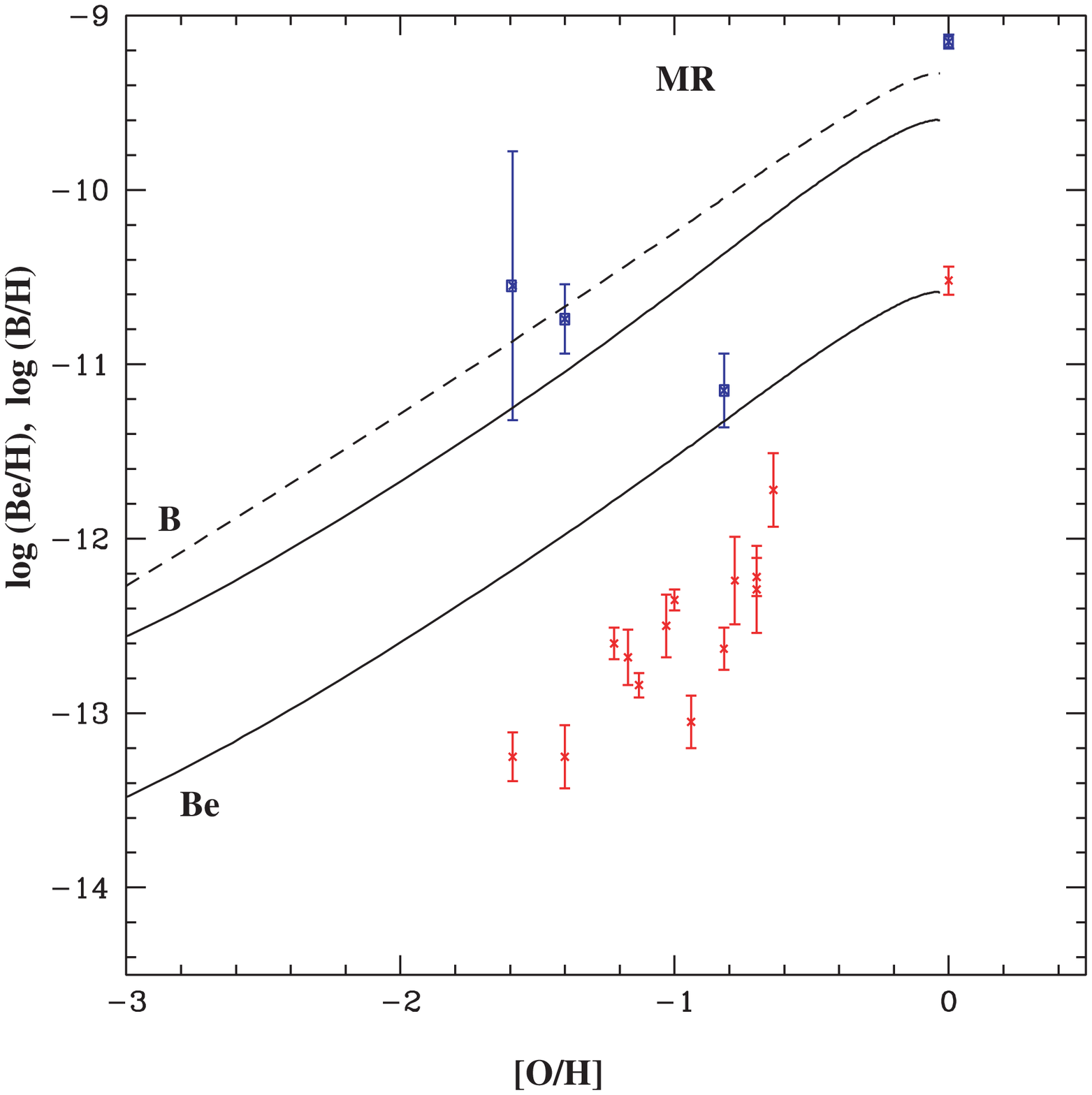, height=3.0in}
\end{center}
\caption{As figure \protect{\ref{beba}} with data and models shown as a function of [O/H].}
\label{beboa}
\end{figure}

As argued in \citep{fo,fovc}, a better way to determine the primary vs secondary nature
of GCRN is to track the evolution of Be relative to O/H.  This is shown in Fig.  \ref{beboa}.
Again, the IRFM data is plotted on the left and MR data on the right.  The full set of 24 points
based on the IRFM \citep{fovc} gave a slope of 1.80 $\pm$ 0.17. 
This slope is more characteristic of secondary GCRN and the cross-over between 
primary and secondary was determined to be at [O/H]$_{eq}$ = -1.88.
 When limited to the 13 points
with MR parameters, this slope was reduced to 1.51 $\pm$ 0.36 with a cross-over 
at [O/H]$_{eq}$ = -1.59.
However, when the MR parameters are used, the slope drops to 1.36 $\pm$ 0.27
and the cross over occurs at [O/H]$_{eq} \approx$ -1.4, and signals a greater
role for primary nucleosynthesis.  
However, as one can see, the models have a difficult time tracking the data
due to the enhanced O/H abundances. 
This is true for the IRFM data and accentuated in the MR data.
In this context, to improve this evolution
it is required to modify oxygen yields in massive stars, specifically at low metallicity.

The models shown in Figs. \ref{beba} and \ref{beboa} were constructed 
using stellar nucleosynthesis yields which produce a flat relation between
[O/Fe] and [Fe/H] at low metallicity. 
 As discussed above,  the data
do not respect this relation and this is the main reason the models
seem to overproduce \be9 when plotted against [O/H].  In fact, they 
are underproducing [O/H] for a given [Fe/H].
Since Fe yields are notoriously more uncertain than O yields, 
we have also constructed a set of models in which
the iron yields in Type II supernovae are reduced relative to 
the values in \citet{ww95}. In these models, the iron yields are set
by the oxygen production so as to fit the O/Fe trend in consideration.

\begin{figure}[ht]
\begin{center}
\epsfig{file=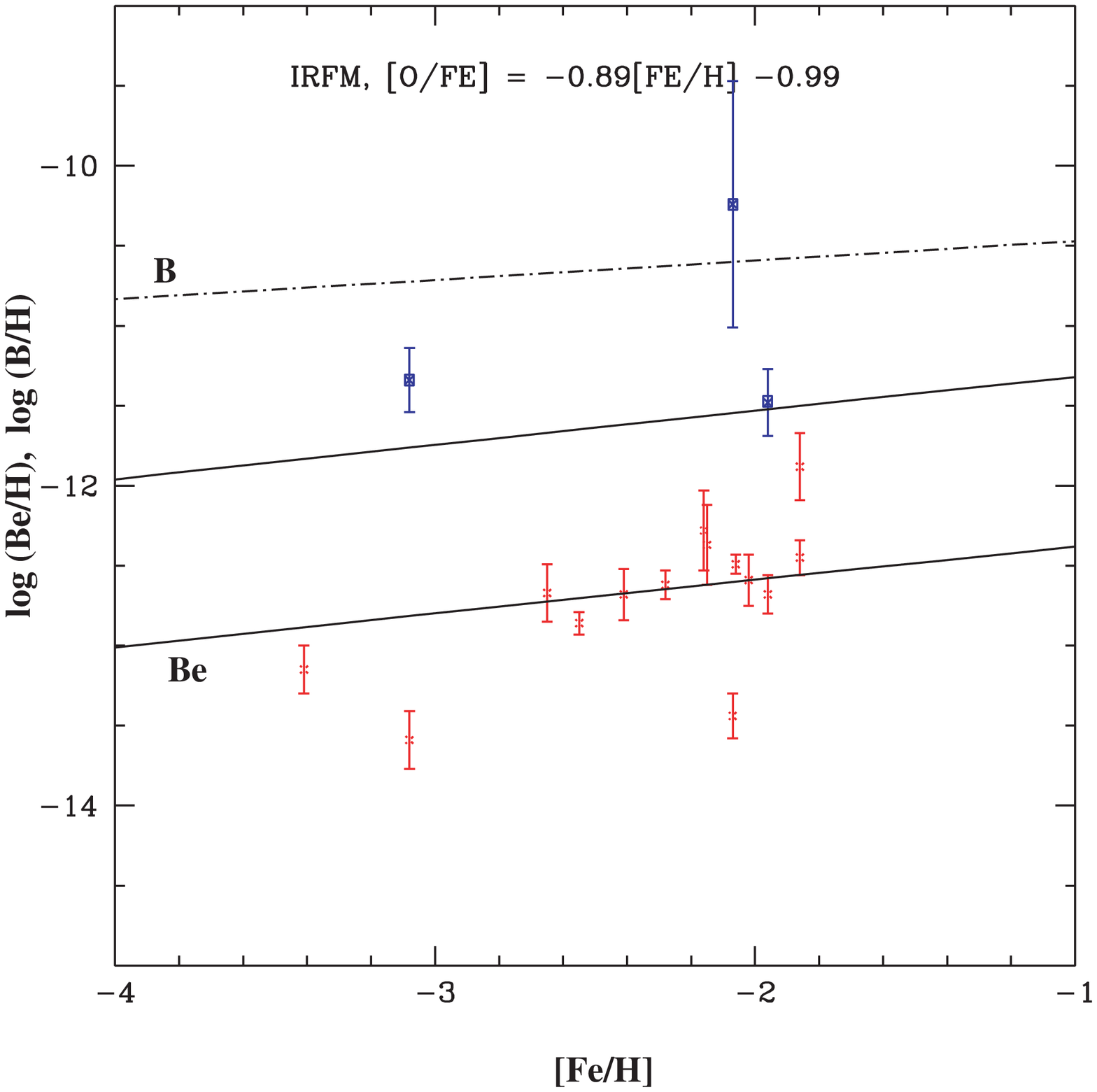, height=3.0in}
\epsfig{file=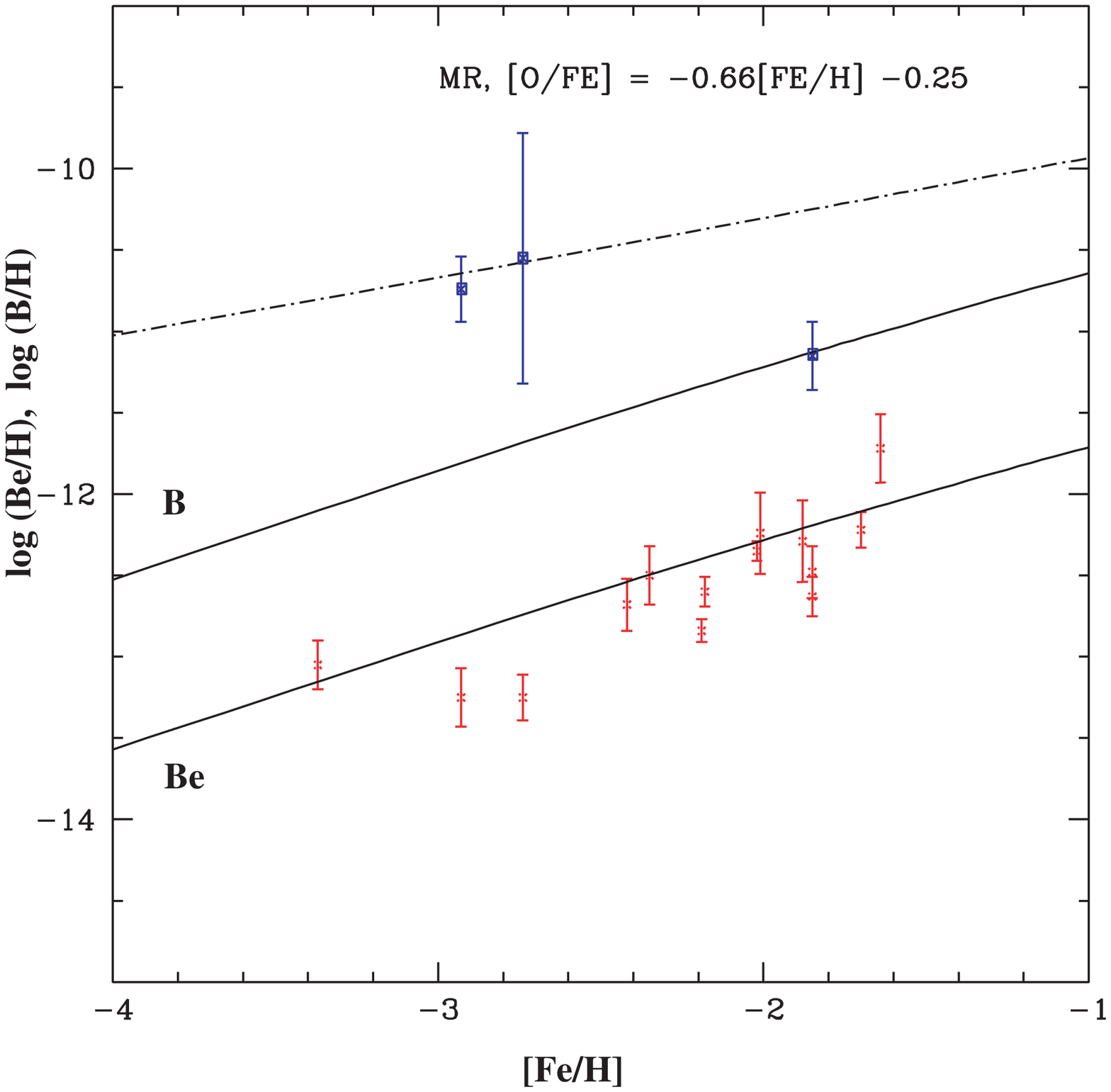, height=3.0in}
\end{center}
\caption{As in Figure \protect{\ref{beba}} but with Fe yields adjusted so 
as to insure the derived relation between [O/Fe] and [Fe/H].}
\label{bebb}
\end{figure}

\begin{figure}[ht]
\begin{center}
\epsfig{file=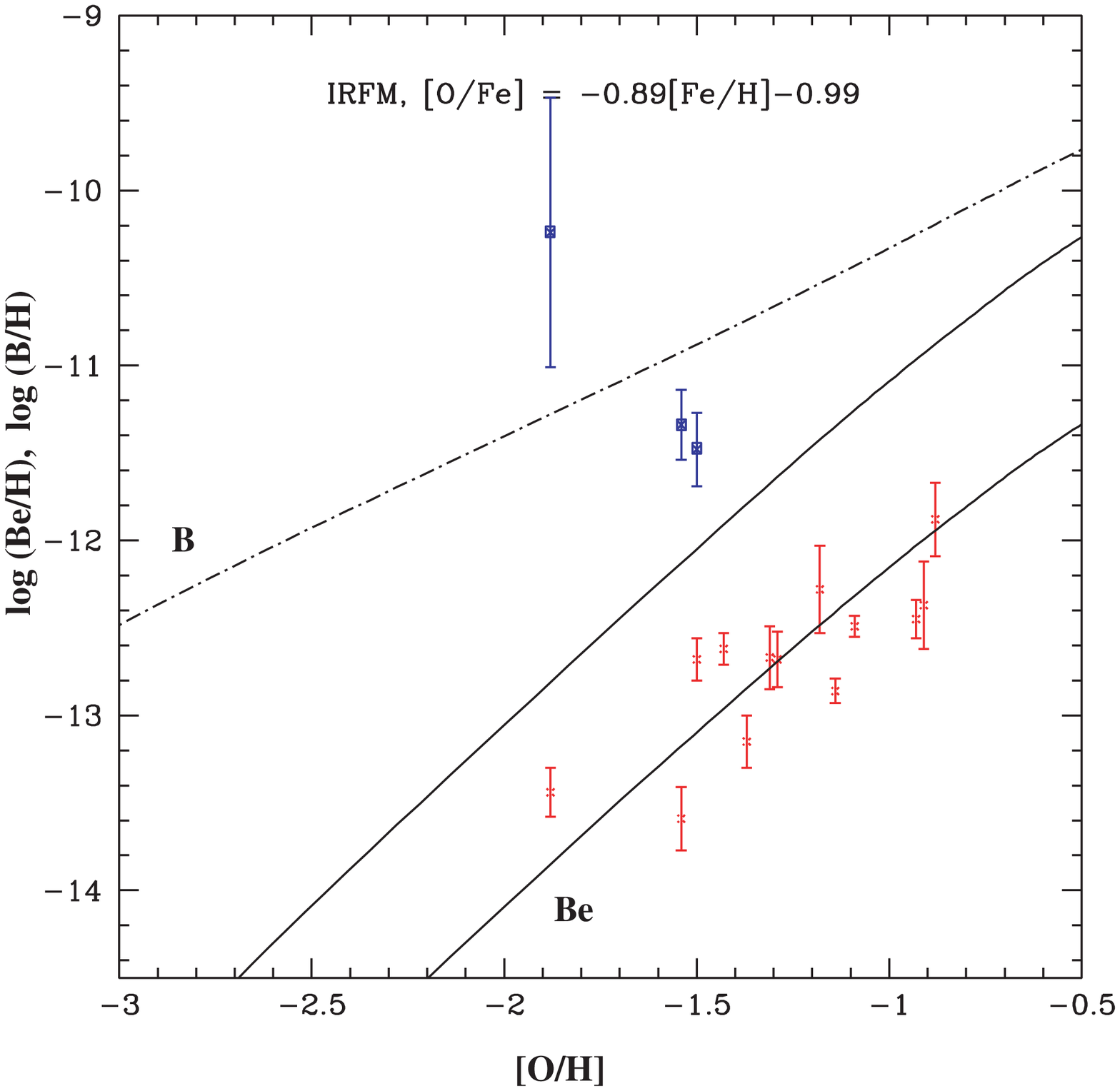, height=3.0in}
\epsfig{file=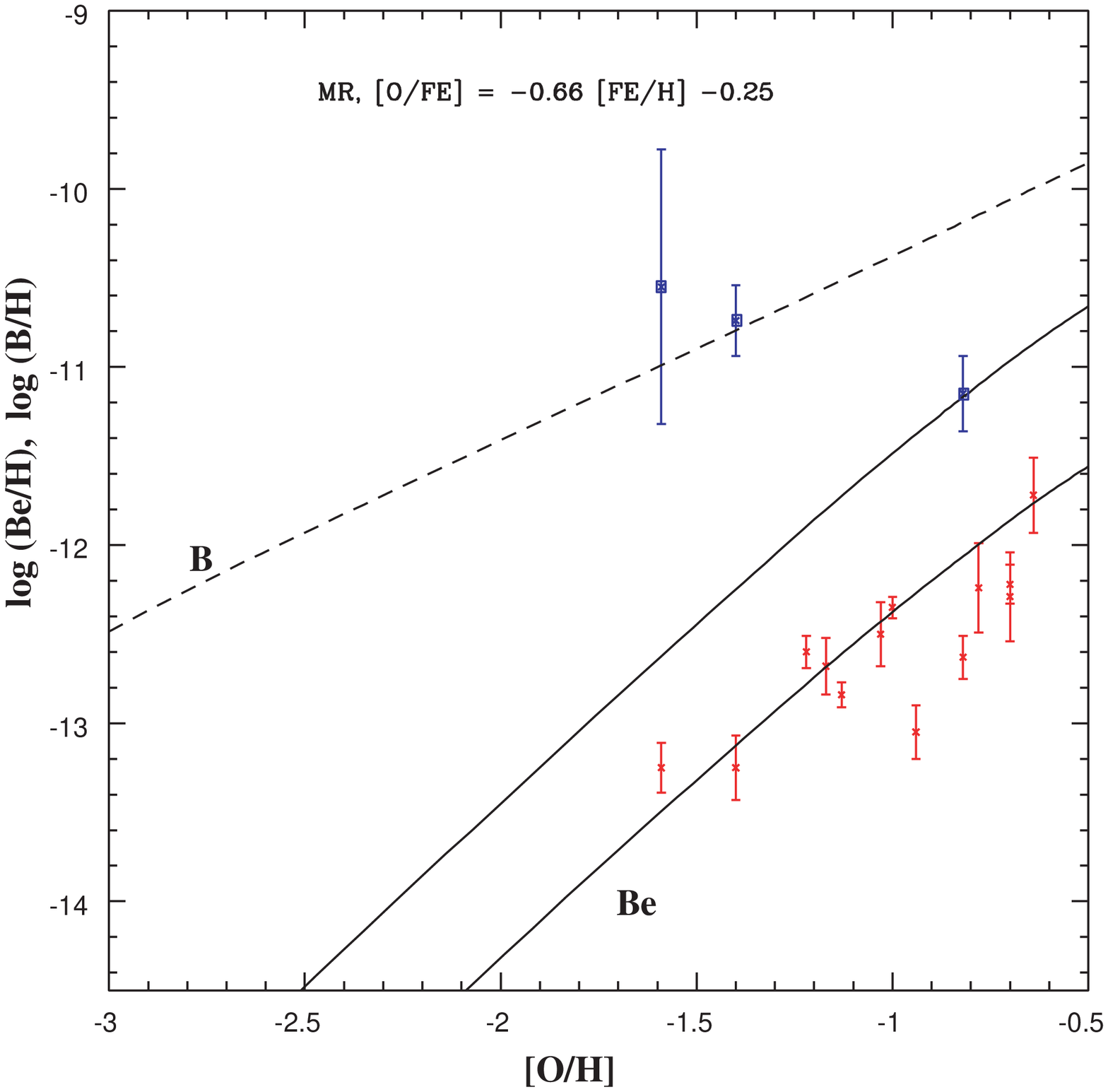, height=3.0in}
\end{center}
\caption{As in Figure \protect{\ref{bebb}} with data and models shown as a function of [O/H].}
\label{bebob}
\end{figure}

In Figs. \ref{bebb} and \ref{bebob} we show the corresponding
models as a function of [Fe/H] and [O/H] respectively.
Using the determined relation between [O/Fe] vs [Fe/H] for each of the data
sets discussed above, we see that standard GCRN can describe
both the Be dependence versus [Fe/H] and [O/H].
Note however, that the employed fits of [O/Fe], are only applicable 
at very low metallicity and can not be extrapolated to [Fe/H] = 0.
Indeed at [Fe/H] = 0,  both the IRFM and MR fits would yield 
subsolar oxygen abundances.  In the case of the IRFM, 
an extended range of data is available (yielding a milder slope for 
[O/Fe] vs [Fe/H]) and as such can be extended to higher metallicity.
In the case of the MR data, it would be very useful to have a more complete
sample of stellar parameters for those stars with Be, B and O abundances
to test both the [O/Fe] vs [Fe/H] relation and the need of a primary source for 
GCRN.

Finally, we comment on the post-BBN production of Li.
Invariably, the increase in Li due to GCRN and the $\nu$-process
will lead to a slope in the relation between log(Li/H) and [Fe/H] which depends on
the primordial value of Li (recall that these quantities are the logs of abundances).
The slope found in the data of \citet{rnb} was marginally consistent with model calculations
as discussed in \citep{rbofn}.  For example, a log slope of 
$d\log {\rm Li}/d\log {\rm Fe} = 0.037 - 0.074$
was found
assuming \li7/H$_p$ = 0.9 -- 1.9 $\times 10^{-10}$.  For small slopes, the log slope is roughly
inversely proportion to the primordial Li abundance.  At the large end of the range,
a slope of 0.074 over a metallicity range of [Fe/H] from -3 to -1 would
correspond to a change in log(Li/H) of about 0.15 dex which is larger than the observational uncertainty.
However, for a primordial abundance of 2.34 $\times 10^{-10}$ the slope drops to  0.03
and for the BBN value of 4.3 $\times 10^{-10}$, it becomes 0.016.  These are both too small to
be observed as the overall increase in Li is the observational error of individual points.
Thus, it is not a surprise that the MR data do not show a slope for Li with respect to [Fe/H].

In the models considered here, we find a slightly larger log slope due in part to the 
enhanced primary component of the LEC model. Our slope is further enhanced 
(relative to \citet{rbofn})
as the $\nu$-process in these models plays a more important role
due to the enhanced B abundances with MR parameters.   In fact, the slope is in fact not
constant, but increases with metallicity.  In the metallicity range, [Fe/H] = -2 to -1, 
we would expect an increase which is large enough to be observable
beyond individual scatter.  Most of the Li data analyzed by \citet{mr}, lies at [Fe/H] $< -1.5$,
so it would be difficult to claim evidence for the enhanced primary component in Li at this time.

\section{Summary}

\li7 plays a key role as a bridge between big bang nucleosynthesis and 
galactic cosmic ray nucleosynthesis.  At present, there is a significant discrepancy
between the BBN-predicted \li7 abundance assuming a baryon density
consistent with the concordance model derived from observations 
of anisotropies in the microwave background, and the abundance
determined from the observations of \li7 in the atmospheres of halo stars.
The discrepancy is large enough ( a factor of 2 -- 3), that it appears
unlikely to be resolved at the level of the nuclear physics inputs to BBN
calculations.  The remaining conventional options (i.e., not invoking physics beyond the 
Standard Model) are an adjustment of the stellar input parameters needed to 
extract a \li7 abundance from observations, or stellar depletion of \li7.
Stellar depletion certainly remains a possibility, though models must be
constructed to avoid dispersion in the \li7 abundances over a wide range of 
stellar parameters.  

Motivated by the recent analysis of \citet{mr} who argue for a revised
temperature scale for low metallicity plateau stars, we have explored
the consequences of the MR temperature scale
on the post BBN production of \li7 in association
with the production of Be and B in galactic cosmic ray nucleosynthesis.
\citet{mr} have argued that the surface temperatures of low metallicity
halo stars have been systematically underestimated.  As a consequence, they
revise upwards the observed abundance of \li7. While this shift in \li7/H does
not completely resolve the problem,  the discrepancy is reduced to less than a factor of 
two and is within 2-$\sigma$ of the BBN prediction.  

In addition to \li7,  both Be and B are also observed in many of the same halo stars.
As such, any revision to the set of stellar parameters used to obtain element 
abundances which affect \li7 will also affect Be and B.  Furthermore, since
the BeB elements are produced in spallation processes which involve CNO, 
the oxygen abundance (used here as a tracer of metallicity) is of interest and is also affected by
the change in stellar parameters. 

In this paper, we have considered the effects of the new temperature scale
proposed by \citet{mr} on the abundances of Be/H, B/H, and O/H. 
While the Be abundances are not overly affected by the shift to higher temperatures,
the B and O abundances are very sensitive to the assumed input temperature.
As such, we find that B abundances and in particular, B/Be abundance ratios
are significantly higher at low metallicity than in previous considerations.  A 
more complete compilation of stellar parameters, particularly for the low metallicity stars
with observed B, will be a key towards distinguishing models of cosmic ray nucleosynthesis.
In particular, models in which a strong primary component based on low energy cosmic
rays found in the vicinity of superbubbles, do not predict enhancements of 
B/Be at low metallicity, in contrast to secondary models of GCRN when combined
with the primary production of B through the $\nu$-process.

Perhaps a more delicate question concerns the O/H abundances found in halo stars. 
In previous surveys (based on OH lines), [O/Fe] abundances were seen to 
rise moderately at low metallicity.  An enhancement of stellar temperatures, particularly
when the enhancement is systematically high at low metallicity would
imply a sharper increase in [O/Fe] at low metallicity.  Standard chemical evolution models based on
published yields of O and Fe (eg. from \citep{ww95}) can not account for the increased 
O/Fe ratio.  This is apparent in our attempts to model the evolution of Be using standard yields.
If high mass and low metallicity stars produce significantly less Fe than their higher metallicity
counter parts, the evolution  of Be and B versus both O/H and Fe/H can be successfully modeled.

Adjusting the stellar input parameters (primarily the surface temperature) has the potential for resolving the \li7 problem. We have shown that such an adjustment has important consequences for Be and B evolution in the early Galaxy,
as well as the O/Fe ratio. 
In particular, a steep O/Fe trend at low metallicities is suggested by the MR temperature scale 
which places demands on Fe and/or O supernova yields,
and in turn challenges models with a strong primary component
of Be as well as supernova and nonthermal
nucleosynthesis models.
However, we believe that this avenue for pursuing the \li7 problem is encouraging, and significant progress can be made by a thorough and systematic reexamination of the existing
Be, B, and O abundances in low metallicity halo stars.

\section*{Acknowledgments}

The work of K.A.O. was partially supported by DOE grant
DE-FG02-94ER-40823. The work of B.D.F. was supported by the National
Science Foundation under grant AST-0092939.  The work of E.V.F. was
partially supported by PICS 1076 USA/CNRS.

%\end{references}

\end{document}